\documentclass{article}

\usepackage[final,nonatbib]{neurips_2021_ml4ps}

\usepackage[utf8]{inputenc} 
\usepackage[T1]{fontenc}    
\usepackage{hyperref}       
\usepackage{url}            
\usepackage{booktabs}       
\usepackage{amsfonts}       
\usepackage{nicefrac}       
\usepackage{microtype}      
\usepackage{xcolor}         
\usepackage{graphicx}
\usepackage{ulem}

\title{Unsupervised topological learning approach\\ of crystal nucleation in pure Tantalum}

\author{%
 S\'ebastien Becker\\
   \texttt{sebastien.becker@univ-grenoble-alpes.fr} \\
   \And
   Emilie Devijver  \\
   \texttt{emilie.devijver@univ-grenoble-alpes.fr} \\
   \AND
   R\'emi Molinier \\
  \texttt{remi.molinier@univ-grenoble-alpes.fr} \\
   \And
   No\"el Jakse \\
   \texttt{noel.jakse@univ-grenoble-alpes.fr} \\
 Universit\'e Grenoble Alpes, CNRS, Grenoble INP, SIMaP, LIG et IF\\
  F-38000 Grenoble, France
}

\begin{document}

\maketitle

\begin{abstract}
 Nucleation phenomena commonly observed in our every day life are of fundamental, technological and societal importance in many areas, but some of their most intimate mechanisms remain however to be unraveled. Crystal nucleation, the early stages where the liquid-to-solid transition occurs upon undercooling, initiates at the atomic level on nanometer length and sub-picoseconds time scales and involves complex multidimensional mechanisms with local symmetry breaking that can hardly be observed experimentally in the very details. To reveal their structural features in simulations without a priori, an unsupervised learning approach founded on topological descriptors loaned from persistent homology concepts is proposed. Applied here to a monatomic metal, namely Tantalum (Ta), it shows that both translational and orientational ordering always come into play simultaneously when homogeneous nucleation starts in regions with low five-fold symmetry.\end{abstract}

	Understanding homogeneous crystal nucleation under deep undercooling conditions remains a formidable issue, as crystallization is essentially heterogeneous in nature and initiated from impurities, surfaces, or near grain boundaries that often hinder its occurrence \cite{Sosso2016}. Unreachable until very recently, experimental observations of early stages of nuclei was achieved by a \textit{tour de force} using time tracking of three-dimensional (3D) Atomic Electron Tomography \cite{Zhou2019} of metallic nanoparticles. Those complex phenomena remain to date out-of-reach experimentally for bulk systems, thus hindering our theoretical understanding. This line of research still belongs mostly to the domain of atomic-level simulations and more particularly to molecular dynamics (MD) with generic interaction models \cite{Auer2001,ten1995}. To reach statistically meaningful events, large scale simulations are required\footnote{This still remains challenging as only few studies are providing now million-atom simulations for monatomic metals \cite{Sosso2016}.}.

	To identify translational and orientational orderings during homogeneous nucleation in MD simulations, an unsupervised learning approach based on topological data analysis (TDA) signatures, and more precisely persistent homology (PH) \cite{Motta2018,Carriere2015} was developed. PH is an intrinsically flexible, yet highly informative, tool which detects meaningful topological features deduced from atomic configurations. It was successfully applied very recently to characterize structural environments in metallic glasses \cite{Hirata2020}, ice \cite{Hong2019} and complex molecular liquids \cite{Sasaki2018}. Always used as a structural analysis in these studies, the originality here is to use PH as a translational and rotational invariant descriptor to encode the local structures required for the clustering method. More precisely, a persistence diagram is drawn from each local structure and then encoded into a topological vector as in \cite{Carriere2015}. Each coordinate of the topological vector is associated to a pair of points $(x,y)$ in a persistence diagram $D$ for a fixed level of homology, except the infinite point, and is calculated by 
	\begin{equation}
		m_D(x,y) = \min \{\|x-y\|_\infty, d_\Delta(x), d_\Delta(y)\},
	\end{equation}
	where $d_\Delta(\cdot)$ denotes the $\ell^\infty$ distance to the diagonal, and those coordinates are sorted by decreasing order. For the clustering, a model-based method is used, namely Gaussian Mixture Models (GMM) \cite[Chapter 14]{tibshirani} and its estimation by an Expectation Maximization (EM) algorithm \cite{Dempster}. The number of clusters is selected by Integrated Criterion Likelihood (ICL, \cite{ICL}), a refinement for clustering of Bayesian Integrated Likelihood (BIC, \cite{BIC}). The inferred model from the method called hereafter TDA-GMM, is used to identify and describe the structural and morphological properties of the nuclei as well as their liquid environment at various steps of  crystal nucleation. 
	\begin{figure}[tb!]\centering
		\includegraphics[width=0.9\textwidth]{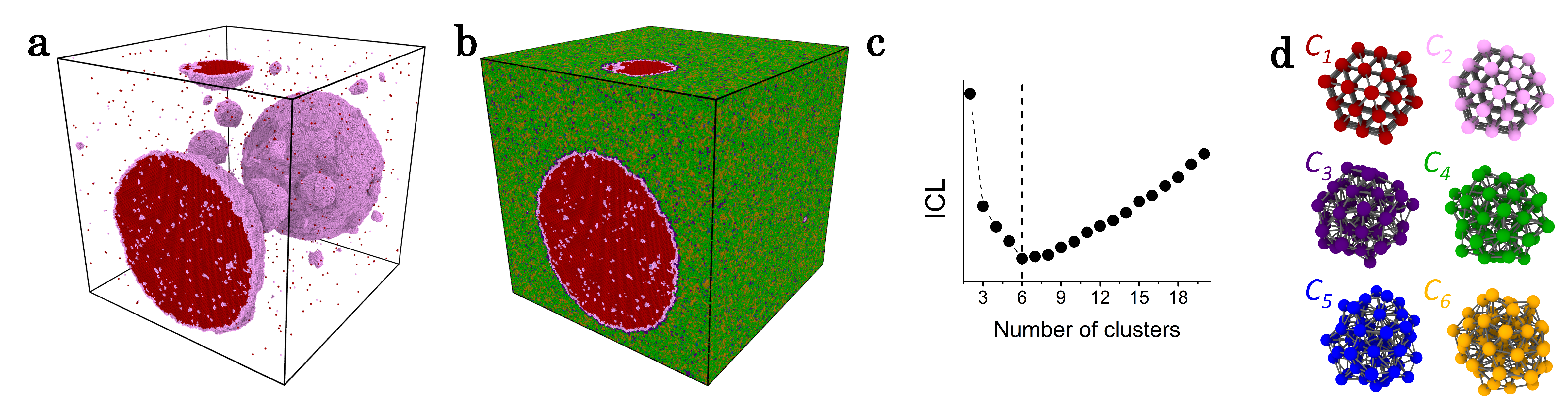}
		\caption{\textbf{Unsupervised learning of homogeneous nucleation.} Snapshot of a ten-million atom MD simulation of Ta during nucleation along the $T=1900$~K isotherm (a and b). Independent local atomic structures form a train set represented in the descriptor space by $215$ PH components up to the second order. (c) Evolution of the ICL criterion as a function of number of clusters is used to get the optimal number of clusters  shown in (d). In (a) the snapshot is represented only with atoms in cluster $C_1$ and cluster $C_2$ revealing all nuclei, while in (b) atoms of all clusters are displayed.}
		\label{fig:Fig1-method}
	\end{figure}
	With this unsupervised approach, the homogeneous nucleation process was studied in liquid Ta, a monatomic metal having an underlying body-centered cubic (bcc) crystalline phase. Large-scale molecular dynamics simulations comprising ten million atoms were performed. Figure \ref{fig:Fig1-method} depicts the result of the methodology. A configuration of the simulation is chosen during crystal nucleation as described below. As it contains many nuclei with different sizes and a substantial amount of liquid, it is considered as representative of the phenomenon. From its inherent structure \cite{Stillinger1982}, a training set of $5$ $000$ non overlapping local spherical structures (encoded trough their topological vectors)  within a cutoff radius of $6.8$ \AA{} was sampled for the unsupervised learning, with the constraints of covering the entire simulation box uniformly and randomly. Among all possible sets upon applying the GMM, the one with $6$ clusters (later on denoted by $C_1$ through $C_6$) shown in \ref{fig:Fig1-method} (d) was automatically inferred to be representative of the system based on the minimum ICL criterion \ref{fig:Fig1-method}(c). The snapshot of the simulation box in Fig. \ref{fig:Fig1-method}(a) displays only local structure from clusters $C_1$ and $C_2$, as they show clearly a crystalline order. They reveal all nuclei as it will be seen below, along with their structure, size and morphology out of the simulation box displayed in Fig. \ref{fig:Fig1-method}(b). From this model, each atom of each configuration generated by the MD simulation can be assigned, when considered with its surrounding local structure, to one of the six clusters (the one with the highest probability). Such a clustering training is performed and shows that more than $99.99$ \% of the structures have a probability to belong to the most probable Gaussian component greater than $0.999$, even for structures not in the initial training set. An analysis of the eigenvalues of the covariance matrices shows elliptical shapes, which proves the necessity of the GMM with general covariance matrices compared to simpler unsupervised algorithms (e.g., $k$-means would only fit hyperspheres).
	
\begin{figure}[tb!]\centering
		\includegraphics[scale=0.125]{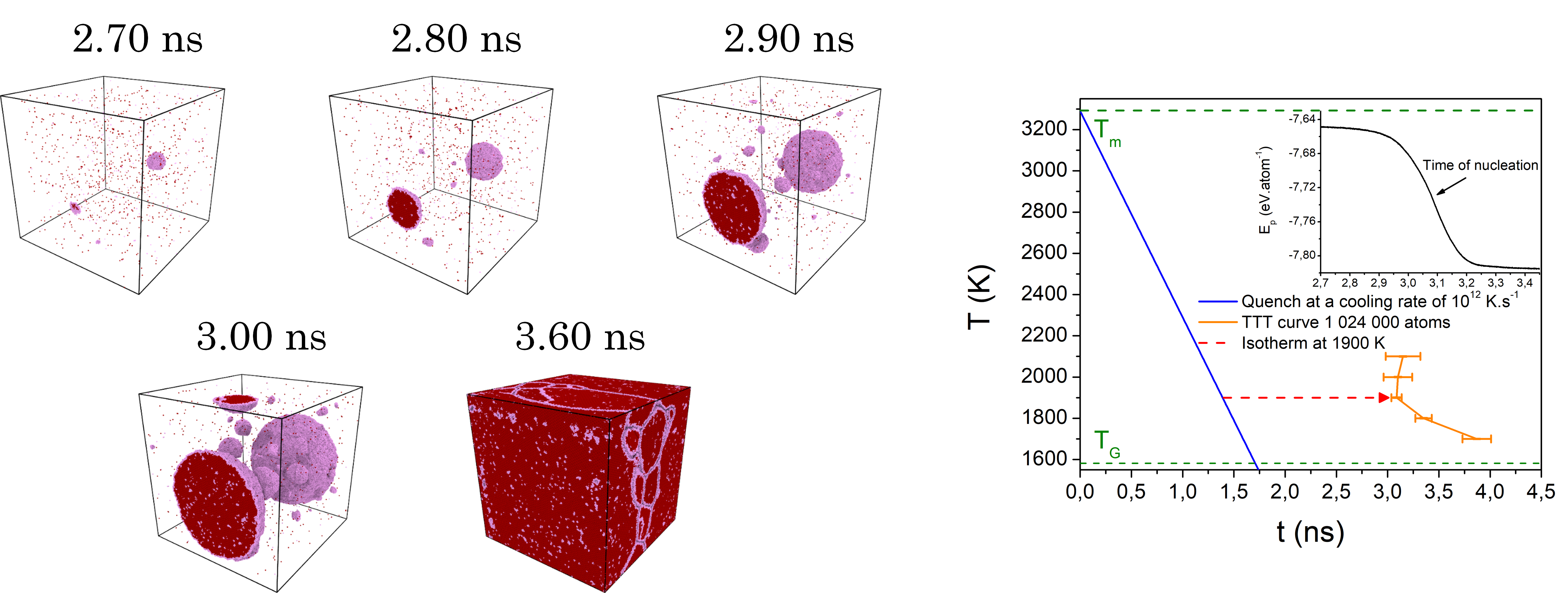} 
		\caption{\textbf{Homogeneous nucleation in Ta undercooled liquids.} Snapshots of the MD simulations, during isothermal nucleation at different times for temperatures close to the nose of the Time-Temperature Transformation (TTT). From stored configurations during fast cooling (blue curves), nucleation events along several isotherms were observed by monitoring the sharp drop of the internal energy (inset). The average nucleation times $\tau_N$ (symbols) were determined from $5$ independent simulations for each temperature giving the TTT curves in the vicinity of the nose (orange lines).}
		\label{fig:Fig2-observation}
	\end{figure}

	Figure \ref{fig:Fig2-observation} shows typical homogeneous nucleation events in undercooled Ta during an isothermal process close to the nose of the time-temperature-transformation (TTT) curve\footnote{which can be done by standard MD simulations without the need of an accelerated method \cite{Allen2009}.}. The liquid above the melting point $T_{M}$ (at $T=3300$ K) was first quenched down at ambient pressure to the glass transition sufficiently rapidly to avoid nucleation. From stored configurations during cooling, the TTT curves in the vicinity of the nose were built from observation of the nucleation along several isotherms as shown in Figs. \ref{fig:Fig2-observation}. An isotherm slightly above the TTT nose is chosen for the analysis ($T=1900$ K). From chosen configurations during the nucleation and growth process, the clustering is obtained from application of the trained model. Strongly growing fraction of mainly two clusters, concomitant to the sharp drop of the energy, is observed. Only local structures belonging to these clusters are drawn in Figures \ref{fig:Fig2-observation}, revealing evidently the nuclei and their evolution in time, recalling that solely the topological vector is describing the local structure. The nuclei morphologies show globular shapes that are rather spherical, characteristic of high $\Delta T$. Interestingly, atoms from one of the two clusters (in red) are mainly located inside the nuclei while atoms from the second one (in pink) steadily remain essentially at the border upon growing. They stay finally at grain boundaries after full solidification of the simulation boxes.
	The vast majority of the embryos\footnote{Nuclei smaller than the critical size of 65 atoms.} seen in Fig. \ref{fig:Fig2-observation} dissolves back to the liquid while those attaining the critical size are rare and grow. The large simulation box allows to follow the nucleation process for a longer time, sufficient to observe more secondary nucleation events \cite{Shibuta2017}. 
		
\begin{figure}[tb!]\centering
		\includegraphics[width=0.75\textwidth]{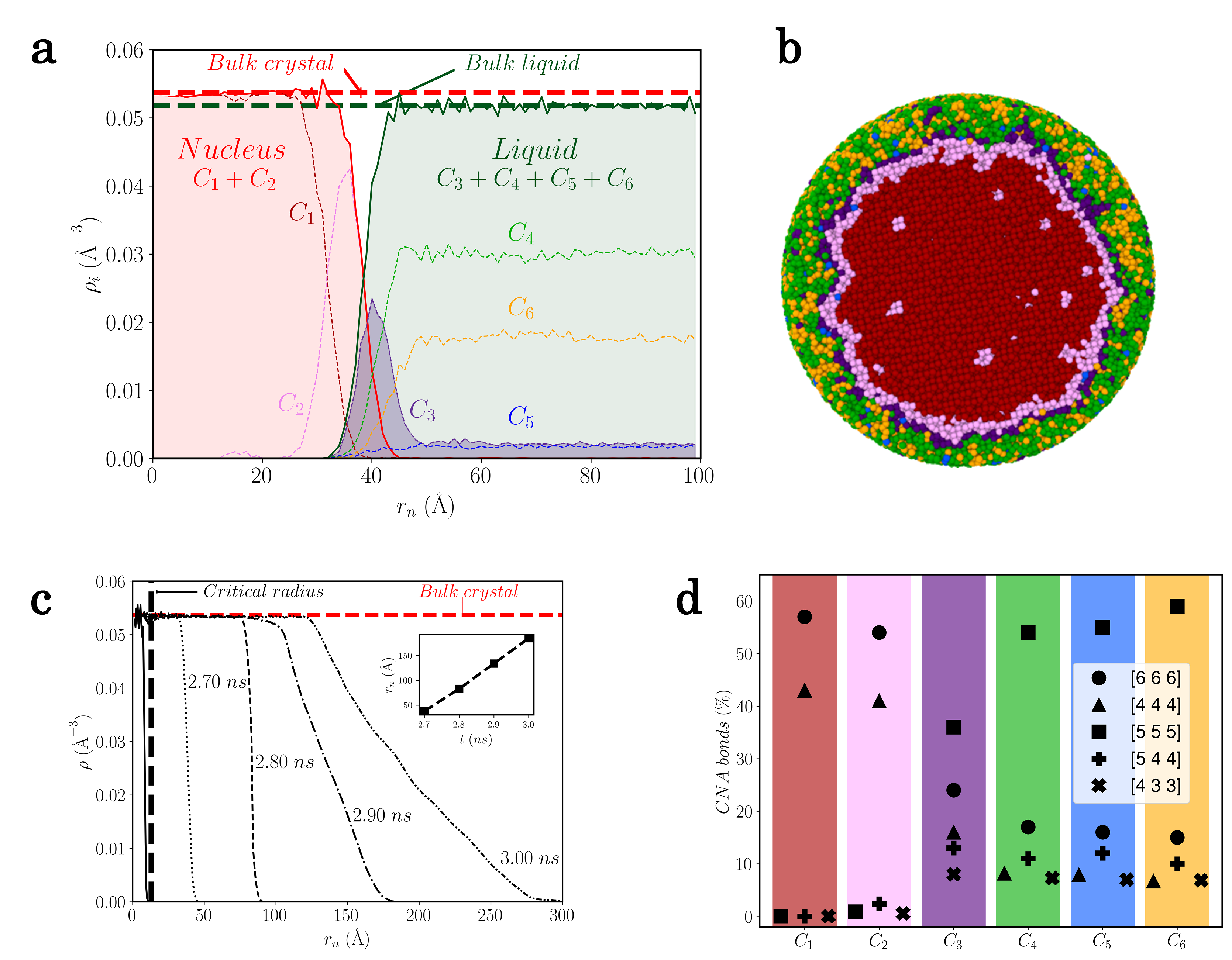}
		\caption{\textbf{Translational and bond-orientational order parameters.} (a) Radial density profile of the largest nucleus during the growth at $2.7$ ns along the $T=1900$ K isotherm. The red and blue dashed horizontal lines correspond respectively to the average bulk crystalline density and average bulk undercooled liquid without nucleation events (b) Corresponding slice of the nucleus through its centre and the surrounding liquid where atoms have been coloured according to the cluster they belong to (see Fig. \ref{fig:Fig1-method}(d)). (c) Total radial density profile of the largest nucleus during growth before solidification. Inset: time evolution of the radius of the nucleus. (d) Bond-orientational order in terms of bonded pairs of the Common-Neighbor Analysis \cite{Faken1994} for each cluster of the model.}
		\label{fig:Fig3-analysis}
	\end{figure}

The nucleation process is characterized at least by two order parameters, the translational order (TO) and the crystalline ordering called hereafter the bond orientational order (BOO). A dedicated representation of the TO is the number density. It is primarily applied to the embryos and the nuclei at different stage of the growth, through the radial partial atomic density profiles $\rho_i(r) = N_i(r)/\frac{4\pi}{3} [(r+\Delta r)^3-r^3]$ as a function of distance $r$ of the estimated center of the nucleus, $N_i(r)$ being the number of atoms belonging to cluster $C_i$ in a spherical shell of radius $r$ and thickness $\Delta r=1$ \AA. Fig. \ref{fig:Fig3-analysis}(a) depicts the density profiles $\rho_i(r)$ for all $6$ clusters for the largest nucleus shown in Fig. \ref{fig:Fig2-observation}(a) and its surrounding liquid at time $2.7$~ns. The corresponding slice of the nucleus through its center is drawn in Fig. \ref{fig:Fig3-analysis}(b). Thus, the nucleus is defined by atoms belonging to clusters $C_1$ and $C_2$ as described above, atoms of $C_1$ forming the center of the nucleus, while atoms of $C_2$ being mainly located at its border. It should be noted that atoms of cluster $C_3$ are mainly located at the boundary of the nucleus, but they cannot be considered as being part of it, as they are also present in the entire box. From the total density profile of the nucleus $\rho_N(r) = \rho_1(r)+\rho_2(r)$, it can be seen clearly that the density of nucleus has already reached at this stage the one of the bulk crystal at the same temperature. Defining the remaining clusters ($C_3$ to $C_6$) as belonging to the liquid yields to a total density profile $\rho_L(r) = \sum_{i=3}^6\rho_i(r)$ showing that even in the vicinity of the nucleus the liquid is negligibly influenced by its presence, keeping the density of the bulk undercooled liquid. 
Fig. \ref{fig:Fig3-analysis}(c) shows the evolution of the density profile $\rho_N(r)$ at different times of the growing process. The average radius $r_N$ of the nucleus is taken as the value of $r$ at half-maximum of $\rho_N(r)$ and its evolution with time is shown in the inset, displaying a linear behaviour in agreement with CNT \cite{Sosso2016}. Whatever the size of the nuclei, the density of the inner part is close to the bulk crystal. More importantly, this is all the more true for all the embryos below the critical size up to a single atomic structure corresponding to the minimal size of about $65$ atoms belonging to cluster $C_1$ or $C_2$ as identified from their local structure.  
The BOO of each cluster is identified through the Common Neighbor Analysis (CNA) \cite{Faken1994}, chosen as a well-known and robust tool. The CNA signature \cite{Jakse2006} given in Fig. \ref{fig:Fig3-analysis}(d) reveals that structures from clusters $C_1$ and $C_2$ possess respectively a perfect and slightly distorted bcc crystalline ordering confirming the above analysis of nucleation and growth in terms of topological descriptors. Local structures from clusters $C_4$, $C_5$ and $C_6$ display various high degrees of five-fold symmetry (FFS) characteristic of the liquid state together with a small but non negligible degree of bcc ordering, while structures from cluster $C_3$ retains both FFS and bcc order in similar proportions. Such a BOO of the four clusters associated to the liquid agrees well with \textit{ab initio} molecular dynamics simulations \cite{Jakse2004} and was interpreted as compatible with the A15 crystalline phase. This analysis highlights and confirms that the TDA-GMM unsupervised learning approach is a powerful method to capture the structural picture in its finest details.

The question whether the onset of nucleation is initiated primarily by translational or by orientational ordering is still open, and was debated during the last decade with a controversy essentially centered on the hard sphere and associated colloidal systems \cite{Berryman2016,Russo2016}. The small emerging embryos at the onset of nucleation, corresponding to one structure of $55$ to $70$ atoms belonging to $C_1$ or $C_2$ with bcc crystalline BOO, show bond lengths of their bcc lattice close to the density of the bulk crystal. This provides evidence  given the size of embryos that can be detected here: translational and bond-orientational orders appear simultaneously and rule out the scenario in which homogeneous nucleation is driven by BOO first \cite{Russo2012} for metallic systems\footnote{which is consistent with the fact that, unlike hard spheres, metallic systems with strong bonding are more energy driven rather than entropy driven systems.}.

The present unsupervised learning approach was shown to be a powerful tool to unravel the atomic scale mechanisms of crystal nucleation in Ta. Other unsupervised methods can retrieve the dissociation between solid and liquid-like structure. For example, a simple Principal Component Analysis discriminates those two states on the first axis, as well as the famous t-SNE \cite{Maaten2008} that represents the points such that liquid related particles are closer . However, there is no clear frontier between them (whereas our clusters are well defined, as given by the a posteriori probabilities), and there is for example no distinction between cluster 3 and 4, although the interpretation is clear. Our results are in line with the emerging idea that heterogeneities which exist in the undercooled liquid \cite{Russo2016} play the foremost role in the onset of nucleation. Nucleation have been indeed found to start in low FFS regions, which is consistent with Frank’s argument \cite{Frank1952}, with translational and orientational ordering taking place simultaneously in emerging embryos. Moreover, embryos as well as nuclei during the growth possess the bulk crystal density driven by the metallic bond length while the surrounding liquid keeps the bulk liquid density in accordance with the classical nucleation theory \cite{Sosso2016}. However, our analysis reveals also some aspects beyond the CNT, such as nuclei having a diffuse interface with the surrounding liquid. This promising methodology more generally opens the door to a deeper and autonomous investigation of atomic level mechanisms in materials science. The nucleation analysis on multicomponent systems is, for example, especially relevant to enhance materials design. Also, it would be interesting to extend the method to learn the time evolution, e.g. through recent generalization of the persistent homology to time series \cite{Ravishanker2021}.

\section*{Acknowledgments}

We first acknowledge the two reviewers from their helpful remarks, that allow us to improve this paper. We acknowledge the CINES and IDRIS under Project No. INP2227/72914, as well as CIMENT/GRICAD for computational resources. This work was performed within the framework of the Centre of Excellence of Multifunctional Architectured Materials “CEMAM”ANR-10-LABX-44-01 funded by the “Investments for the Future” Program. This work has been partially supported by MIAI@Grenoble Alpes (ANR-19-P3IA-0003). Fruitful discussions within the French collaborative
network in high-temperature thermodynamics GDR CNRS3584 (TherMatHT) are also acknowledged.

\end{document}